\documentclass{ws-procs9x6clean}

\begin{document}

%%%%%%%%%%%%%%%%%%%%%%%%%%%%%%%%%%%%%%%%%%%%%%%%%%%%%%%%%%%%%% 
% title, author(s) and address(es) put here:                 %
%%%%%%%%%%%%%%%%%%%%%%%%%%%%%%%%%%%%%%%%%%%%%%%%%%%%%%%%%%%%%% 
\def\preprintNumber{SLAC--PUB--10319, UCLA/04/TEP/2, CEA--SPhT--T04/012, %
NSF-KITP-04-16}

\title{Cross-Order Relations in ${\mathcal N}=4$ Supersymmetric Gauge Theories}

\author{C.~Anastasiou \lowercase{and} L.~J.~Dixon}
\address{Stanford Linear Accelerator Center, Stanford University, 
         Stanford, CA 94309%\\{\eighttt babis@slac.stanford.edu}
         }

\author{Z.~Bern}
\address{Department of Physics and Astronomy, UCLA, Los Angeles, CA
90095-1547%\\ {\eighttt bern@physics.ucla.edu}
}  

%\author{L.~Dixon}
%\address{Stanford Linear Accelerator Center, Stanford University, 
%         Stanford, CA 94309\\
%         {\eighttt lance@slac.stanford.edu}}

\author{D.~A.~Kosower\footnote{\uppercase{T}alk presented at 
the
3rd \uppercase{I}nternational \uppercase{S}ymposium
on \uppercase{Q}uantum \uppercase{T}heory
and \uppercase{S}ymmetries,
\uppercase{C}incinnati, \uppercase{OH}, \uppercase{S}ept.~10--14, 2003}}
\address{Service de Physique, CEA--Saclay, 
          F--91191 Gif-sur-Yvette cedex, France
         %\\{\eighttt kosower@spht.saclay.cea.fr}
         }


\maketitle

\abstracts{
The anti-de~Sitter/conformal field theory duality conjecture 
raises the question of how the perturbative expansion
in the conformal field theory can resum to a
simple function.  We exhibit a relation between 
the one-loop and two-loop amplitudes whose generalization
to higher-point and higher-loop amplitudes
would answer this question.  We also provide evidence for
the first of these generalizations.
}

%%%%%%%%%%%%%%%%%%%%%%%%%%%%%%%%%%%%%%%%%%%%%%%%%%%%%%%%%%%%%
% The main text of your paper                               %
%%%%%%%%%%%%%%%%%%%%%%%%%%%%%%%%%%%%%%%%%%%%%%%%%%%%%%%%%%%%%

\section{Introduction}

Thirty years ago, shortly after the discovery of asymptotic freedom,
't~Hooft\cite{tHooft} studied the large-order behavior of gauge theories.  He
used the double-line notation, studying the theory in the limit of a
large number of colors $N_c$.  In this limit, the insertion of a gluon
in a diagram, corresponding to increasing the number of loops by one,
costs a power of $g^2$ but gains a power of $N_c$.  It is thus natural to
study the theory in the regime where $g^2 N_c$ is held fixed.  If this
coupling is of order unity, higher-loop contributions are just as
important as lower-order ones.  We can think of this as filling in a
mesh for any given process, so that we obtain a string worldsheet.

This connection was formulated into a duality conjecture by 
Maldacena\cite{Maldacena}.  
In its most studied form, the anti-de~Sitter/conformal field
theory conjecture posits a duality between ${\mathcal N}=4$ supersymmetric gauge
theory (the conformal field theory) at strong coupling and large
number of colors, and the appropriate limit of string theory
(perturbative supergravity) on an anti-de~Sitter background.  This
duality has passed numerous tests\cite{MaldacenaChecks}
 on quantities protected by
supersymmetry, and more recently\cite{UnprotectedChecks} 
on unprotected quantities allowing
another expansion parameter (the large-`spin' limit of BMN\cite{BMN}
operators).

From the viewpoint of 't~Hooft's original work, the AdS/CFT duality
gives rise to another puzzle: how can the entirety of the complicated
perturbative expansion of a quantum field theory be simple enough 
to be expressed at
fixed order in another field theory?  The computation we summarize here
suggests that part of the explanation may lie in special, unexpected
relations between different orders of 
perturbation theory in the ${\mathcal N}=4$
gauge theory.  Eden et al\rlap{.}{\cite{Schubert}}{} have 
alluded to such relationships.

The operator Green functions that have been studied in 
ref.~\refcite{MaldacenaChecks} are
off shell.  In contrast to our intuition that
manifestly Lorentz- and supersymmetry-invariant
calculations should be much harder with increasing numbers of
supersymmetries, the opposite is in fact true for on-shell quantities, in
particular scattering amplitudes.  Indeed, the 
${\mathcal N}=4$ calculations\cite{GSB,Neq4,Fusing,BRY,GravityConnection}
have been pushed further earlier than the corresponding ones in QCD.  Such
calculations thus allow us to probe further into the perturbative
expansion.

\newcommand{\e}{\epsilon}
Most of these calculations were done using the unitarity-based
method first described in ref.~\refcite{Fusing}.  
To compute a one-loop amplitude, one computes tree amplitudes corresponding
to the various cuts of the desired amplitude.  In general, one
must do these calculations with the cut-crossing legs taken in $D=4-2\e$
dimensions, in order to avoid subtraction ambiguities\cite{vanNeerven} 
in the reconstruction
of the full loop amplitude from its cuts.  
At
one loop, only two-particle cuts arise.
For higher-loop amplitudes, we also compute their various cuts.
The amplitudes on either side of the cut now include loop amplitudes,
and for an $l$-loop amplitude, we must consider up to $(l+1)$-particle
cuts.  In practice, it suffices to compute the cuts of these amplitudes
as well, which are again tree amplitudes.  
For the four-point function at two loops, for example, we
have to consider three-particle cuts, with five-point tree amplitudes
on either side of the cut, and two-particle ``double cuts'', each involving
a product of three tree-level four-point amplitudes.  The construction
of the cuts for the ${\mathcal N}=4$ amplitude is surprisingly 
simple\rlap{.}{\cite{BRY}}

Combining the cuts yields the integrand of a Feynman integral for the
entire amplitude.  In order to complete such a calculation, we also
need to perform the integrals.  At one loop, this is a straightforward
task, and a general method for doing so in dimensional regularization
was written down a decade ago\rlap{.}{\cite{Pentagon}}  At two loops, the
inability to compute multi-scale integrals was a barrier to completing
such calculations.  Thanks to recent work on two-loop 
integrals\cite{SmirnovDoubleBox,Tausk}
and general methods
of tensor reductions at higher 
loops\rlap{,}{\cite{IntegrationTechniques}}
the barrier has been 
surmounted.
Many two-loop amplitudes relevant to experiments have since
been computed\rlap{.}{\cite{TwoLoopFourPoint}}
\iffalse
This in turn has allowed
the computation of a long list of two-loop amplitudes of relevance to
experiment\cite{}.  
\fi

\def\tree{{\rm tree\vphantom{p}}}

\iffalse %Mangled display at LANL
The leading-color (planar) contributions to the two-loop amplitude in
the ${\mathcal N}=4$ theory are given entirely\cite{BRY} in terms of the
planar double box\rlap{,}{\cite{SmirnovDoubleBox}}
\vskip 5pt
\begin{equation}
-s t A_4^\tree \Biggl\{
s\hskip -0.2truein%
\vtop{\vskip -0.25truein\epsfxsize=1.0truein\epsfbox{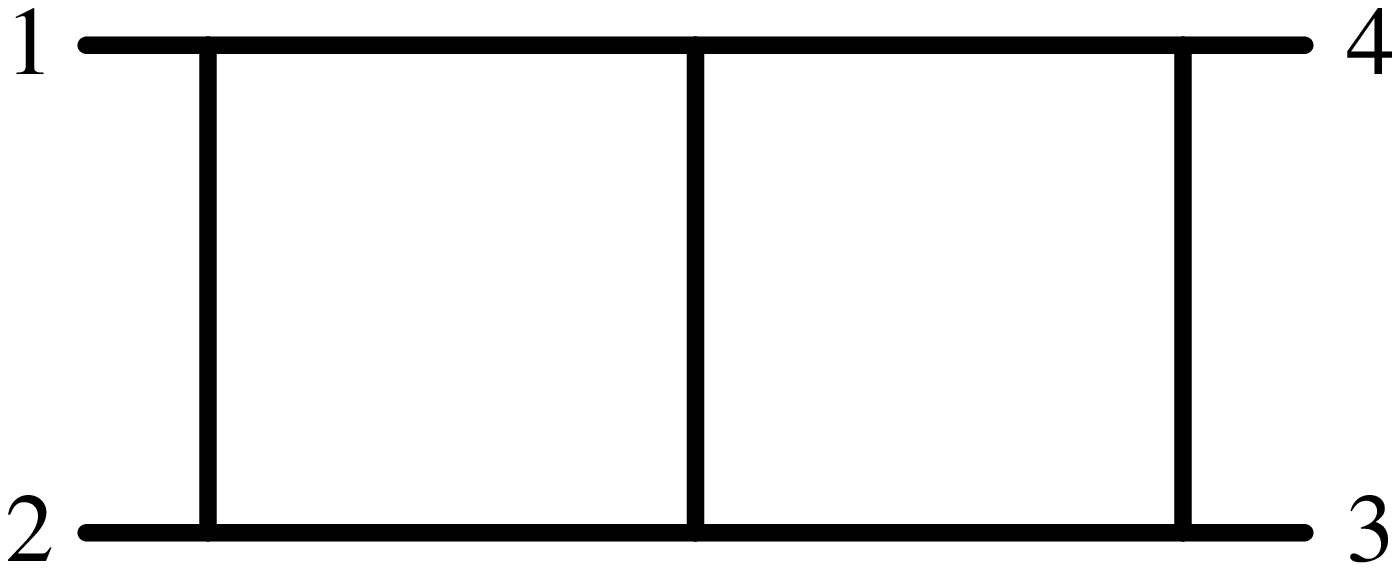}}
\hskip -3.3truein
+t\hskip -0.2truein%
\vtop{\vskip -0.4truein\epsfxsize=0.65truein\epsfbox{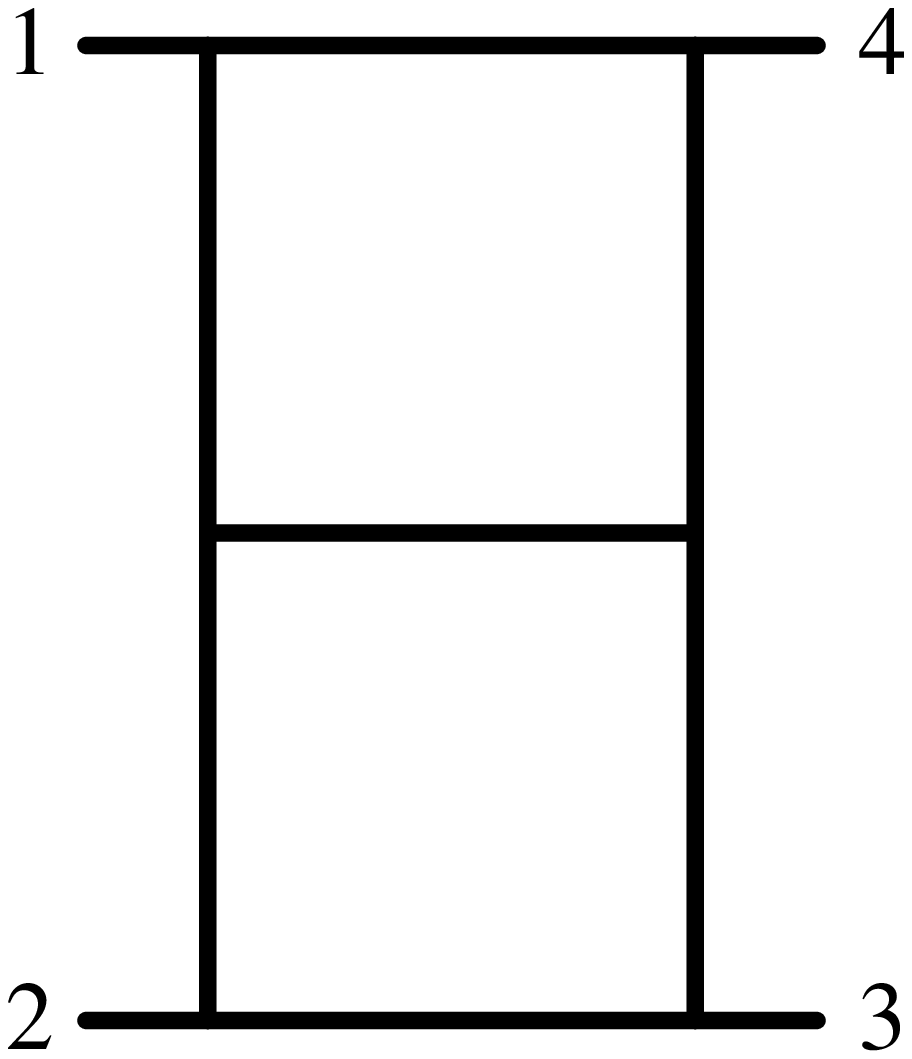}}
\hskip -3.6truein
\Biggl\}\,.
\label{FourPointAmplitude}
\end{equation}
(The ${\mathcal N}=4$ supersymmetry relates all non-vanishing four-point
amplitudes to each other order by order.)
\else
The leading-color (planar) contributions to the two-loop amplitude in
the ${\mathcal N}=4$ theory are given entirely\cite{BRY} in terms of the
planar double box\rlap{,}{\cite{SmirnovDoubleBox}}
\vskip 5pt
\begin{equation}
\hskip -0.3truein
-s t A_4^\tree \Biggl\{
s\hskip 0.1truein%
\vtop{\vskip -0.25truein\epsfxsize=1.0truein\epsfbox{twoloop1.eps}}
\hskip 0.2truein
+t\hskip 0.1truein%
\vtop{\vskip -0.4truein\epsfxsize=0.65truein\epsfbox{twoloop2.eps}}
\hskip 0.1truein
\Biggl\}\,.
\label{FourPointAmplitude}
\end{equation}
\fi

\def\Lloop{{(L)}}
\def\lloop{{(l)}}
\def\oneloop{{(1)}}
\def\twoloop{{(2)}}
\def\Ord{{\mathcal O}}
Define the ratio
$M_n^\Lloop(\e) \equiv A_n^\Lloop/A_n^\tree$
of the $l$-loop leading-color amplitude to the
tree-level one.  The explicit computation from eqn.~(\ref{FourPointAmplitude})
then reveals that
\begin{equation}
M_4^{\twoloop}(\e) =  \frac{1}{2} \Bigl(M_4^{\oneloop}(\e) \Bigr)^2
             + f(\e) \, M_4^{\oneloop}(2\e) - \frac{5}{4} \zeta_4 
             + \Ord(\e)\,,
\label{BasicRelation}
\end{equation}
where $f(\e) \equiv (\psi(1-\e)-\psi(1))/\e
= - (\zeta_2 + \zeta_3 \e + \zeta_4 \e^2 + \cdots)$.
  The relation between different loop orders expressed in this
equation requires the use of polylogarithmic identities,
and involves a non-trivial cancellation of terms between the two
contributing planar double-box integrals.  
Note that terms through $\Ord(\e^2)$ in the one-loop amplitude
contribute to the $\Ord(\e^0)$ terms on the right-hand side,
since they can multiply the $1/\e^2$ terms.

We may naturally expect a similar relation to hold between the $n$-point
amplitudes $M_n^{\twoloop}$ and $M_n^{\oneloop}$.  We can test such
a generalization by studying the collinear limits of the amplitudes.
The behavior of gauge-theory amplitudes in such limits
may be described by splitting amplitudes whose ratio to the tree-level
splitting amplitude we denote by $r_S^{\Lloop}$.   The one- and two-loop
amplitudes factorize as follows\rlap{,}{\cite{OneLoopSplitting}}
\begin{eqnarray}
M_n^{\oneloop}(\e) &\rightarrow& M_{n-1}^{\oneloop}(\e) + r_S^{(1)}(\e) \,,
\label{OneLoopCollinear}\\
M_n^{\twoloop}(\e) & \rightarrow & M_{n-1}^{\twoloop}(\e)
+ r_S^{(1)}(\e) M_{n-1}^{\oneloop}(\e) + r_S^{(2)}(\e) \,. \hskip .5 cm 
\label{TwoLoopCollinear}
\end{eqnarray}

For the conjecture~(\ref{BasicRelation}) to hold for $n$-point amplitudes
as well, $r_S^{\twoloop}$ must be related to $r_S^{\oneloop}$.  To see
what relation is required, we can use the $n$-point analog of 
eqn.~(\ref{BasicRelation}) to rewrite the two-loop quantities in 
eqn.~(\ref{TwoLoopCollinear}) in terms of one-loop quantities and then use
the one-loop collinear limit~(\ref{OneLoopCollinear}).  
Using the method of ref.~\refcite{SplittingApproach},
we find\cite{TwoLoopSplitting} that
\begin{equation}
r_S^{(2)}(\e) =
      \frac{1}{2}\bigl(r_S^{(1)}\bigr)^2 
      + f(\e) r_S^{(1)}(2\e),
\end{equation}
which is exactly what is required for the generalization 
 of eqn.~(\ref{BasicRelation}) to hold.
This does not prove it,
because of possible
 terms which are finite in the collinear limits; but it provides
strong evidence that it is correct.

The form of the ${\mathcal N}=4$ four-point
integrand is known at three loops\cite{BRY}; only
two integrals are needed to complete the computation.  One of these
has been computed recently\cite{SmirnovTripleBox} through $\Ord(\e^0)$.
The  relative
simplicity and regular structure of the integrand
suggest that a similar identity to eqn.~(\ref{BasicRelation})
may hold at higher loops.  Another indication comes from the 
structure of infrared-singular terms, as given by Catani\cite{CataniDiv} at
two loops and Sterman and Tejeda-Yeomans\cite{STY} at three loops.
The structure at two loops is an excellent guide to full 
relation~(\ref{BasicRelation}) between
one- and two-loop amplitudes.  If the same holds true at higher
loops, we expect a relation of the form,
\begin{equation}
M_4^{\Lloop}(\e;s,t) = \frac{1}{L!} 
   \bigl[ M_4^{\oneloop}(\e;s,t)\bigr]^L
 +{\rm lower\ powers\ of\ } M_4^{\oneloop}(m\e;s,t),
\end{equation}
where the form and coefficient
of the leading term are essentially determined by leading-log
resummation.

We can also use the known structure of the infrared-singular terms, which
must cancel against real emission contributions in any `physical' 
quantity, to isolate the remaining finite terms; for these, 
the two-loop relation based on eqn.~(\ref{BasicRelation}) is 
%\begin{equation}
$F_4^\twoloop = 
\frac{1}{2} \bigl[ F_4^\oneloop \bigr]^2 - \zeta_2 \, F_4^\oneloop
            - \frac{21}{8} \zeta_4 \,$,
%\label{TwoloopOneloopFinite}
%\end{equation}
and a higher-loop
relation would then take the form,
\begin{equation}
F_4^{\Lloop}(s,t) = \mathop{\rm Polynomial}\nolimits(F_4^{\oneloop}(s,t))
 = \frac{1}{L!} \bigl[ F_4^{\oneloop}(s,t)\bigr]^L
                     +{\rm lower\ powers}.
\end{equation}

It is worth noting that eqn.~(\ref{BasicRelation}) does {\it not\/}
hold beyond order $\e^0$.  That is, the relation only holds as
$D\rightarrow 4$, where the theory is conformal.  The relation also
appears to be special to the planar or leading-$N_c$ contribution,
as expected from the Maldacena conjecture.

Several open questions merit further study.  It would be desirable,
and is feasible, to compute the two-loop five-point amplitude explicitly
in order to verify the generalization of eqn.~(\ref{BasicRelation}).
Likewise, knowledge of the three-loop four-point amplitude would 
clarify the structure of higher-loop generalizations.  It would be
interesting to identify a symmetry, presumably related to superconformal
invariance, underlying the relation.  Witten's new 
formulation\cite{WittenTopological}
in terms of topological string theory may offer new insight.

Finally, it would be nice
to connect the on-shell ${\mathcal N}=4$ amplitudes 
to off-shell investigations of the duality conjecture.  In this respect,
it is worth noting that the tree-level collinear splitting amplitudes
are closely related to the leading-order Altarelli--Parisi kernel, whose
Mellin moments are the one-loop anomalous dimensions of leading-twist
operators.  Similarly, the next-to-leading order Altarelli--Parisi kernel
can be computed\cite{NLOAP} from the one-loop splitting amplitudes (and other
ingredients), and we expect that the two-loop splitting amplitude can be
used to compute next-to-next-to-leading order anomalous dimensions
in the ${\mathcal N}=4$ gauge theory.

There are also implications for the higher-loop
structure of ${\mathcal N}=8$ supergravity.  
Against prior expectations, this theory was 
argued\cite{GravityConnection}
to be ultraviolet
finite through four loops, based on
the cut-by-cut absence of ultraviolet divergences in the four-point
amplitude.  At five loops, finiteness would require non-trivial cancellations
between different contributions.  The work outlined in this talk shows that
such cancellations occur in maximally supersymmetric gauge theories.
Given the close connection between the integrands of maximally 
supersymmetric gauge and gravity theories, as used in 
ref.~\refcite{GravityConnection}, it is quite possible that 
non-trivial cancellations would take place in the gravitational theory.

%%%%%%%%%%%%%%%%%%%%%%%%%%%%%%%%%%%%%%%%%%%%%%%%%%%%%%%%%%%%%
% Doing Acknowledgement                                     %
%%%%%%%%%%%%%%%%%%%%%%%%%%%%%%%%%%%%%%%%%%%%%%%%%%%%%%%%%%%%%

\section*{Acknowledgments}

We thank M.~Staudacher  for 
helpful discussions, and the Kavli Institute of Theoretical Physics, 
Santa Barbara, for its hospitality.  This research was partially supported
by the US Department of
Energy under contracts DE-FG03-91ER40662 and DE-AC03-76SF00515,
and partially by the National Science
Foundation under Grant No. PHY99-07949.

%%%%%%%%%%%%%%%%%%%%%%%%%%%%%%%%%%%%%%%%%%%%%%%%%%%%%%%%%%%%%
% Doing references:                                         %
%%%%%%%%%%%%%%%%%%%%%%%%%%%%%%%%%%%%%%%%%%%%%%%%%%%%%%%%%%%%%

\end{document}